\documentclass[9pt,preprint2,natbib209]{aastex}

\begin{document}

\shorttitle{Donors of PLMXB} \shortauthors{Zhu et al.}

\title{Donors of Persistent Neutron-star Low-mass X-ray Binaries}
\author{Chunhua Zhu\altaffilmark{1,2}, Guoliang L\"{u}\altaffilmark{2},
 Zhaojun Wang\altaffilmark{2}, Na Wang\altaffilmark{1}}
\email{$^\dagger$chunhuazhu@sina.cn;na.wang@xao.ac.cn}

\altaffiltext{1}{National Astronomical Observatories / Xinjiang
Observatory, the Chinese Academy of Sciences, Urumqi, 830011, China}
\altaffiltext{2}{School of Physical Science and Technology, Xinjiang
University, Urumqi, 830046, China.}

\begin{abstract}
Properties of X-ray luminosities in low-mass X-ray binaries (LMXBs)
mainly depend on donors. We have carried out a detailed study of
donors in  persistent neutron-star LMXBs (PLMXBs) by means of a
population synthesis code. PLMXBs with different donors have
different formation channels. Our numerical simulations show that
more than 90\% of PLMXBs have main sequence (MS) donors, and PLMXBs
with red giant (RG) donors via stellar wind (Wind) are negligible.
In our model, most of neutron stars (NSs) in PLMXBs with
hydrogen-rich donors form via core-collapse supernovae, while more
than 90\% of NSs in PLMXBs with naked helium star (He) donors or
white dwarf (WD) donors form via an evolution-induced collapse via
helium star ($1.4 \leq M_{\rm He}/M_\odot \leq 2.5$) or an
accretion-induced collapses for an accreting ONeMg WD.

PLMXBs with different donors have different properties. In PLMXBs
with MS donors, the orbital periods are between $\sim$ 1 hour and
100 hours, and the mass transfer is driven by donor evolution or
magnetic braking. Our population synthesis code shows that their
X-ray luminosities mainly are around $\sim 10^{36}$ erg s$^{-1}$.
Similarly, in PLMXBs with RG donors via Roche lobe overflow(Roche),
the mass transfer is driven by donor evolution, but orbital periods
are between $\sim$ 10 hours and 1000 hours. Their X-ray luminosities
are around $\sim 10^{37}$ erg s$^{-1}$. The 2 known LMXBs ( Cyg X-2
and GX 13+1) can belong to PLMXBs with RG (Roche). PLMXBs with RG
(Wind) donors have the longest orbital periods and low X-ray
luminosities ($\sim 10^{33}$ erg s$^{-1}$). Their contributions to
X-ray luminosities can be negligible. In PLMXBs with He donors, the
orbital periods are shorter than 80 minutes, and the mass transfer
is mainly driven by magnetic braking. Results of our numerical
simulations predict that PLMXBs with X-ray luminosities around $\sim
10^{38}$ erg s$^{-1}$ mainly come from binaries with He donors. In
PLMXBs with WD donors, the orbital periods are shorter than 1 hour,
and the mass transfer is mainly driven by gravitational radiation.
According to results of our population synthesis code, their X-ray
luminosities are between $\sim 6\times 10^{35}$---$10^{39}$ erg
s$^{-1}$, and most of LMXBs with WD donors are transient.

\end{abstract}

\begin{keywords}binaries: close---pulsar: general---stars:
neutron---X-ray: stars
\end{keywords}
\section{Introduction}
\label{sec:introduction} Low-mass X-ray binaries (LMXBs) were
discovered nearly 50 yr ago, and there are now  $\sim$ 200 known in
the Galaxy \citep{Liu2007}. LMXB is a mass-transferring binary
system with a compact object accretor(a black hole or a neutron star
(NS)), and a low-mass ($\leq 1 M_\odot$) donor. The X-ray luminosity
function is an important characteristic of the LMXBs, and has become
a key tool for studying LMXBs.

Using results of {\it Chandra} observations of old stellar systems
in 11 nearby galaxies of various morphological types and the census
of LMXBs in the Galaxy, \cite{Gilfanov2004} suggested that the total
number of LMXBs and their combined luminosity are proportional to
the stellar mass of the host galaxy. \cite{Postnov2005} suggested
that the flattening of the LMXBs luminosity function at lower than
$2\times10^{37}$ erg s$^{-1}$ might correspond to the transition
from the magnetic stellar wind braking to the gravitational wave
braking mechanism. \cite{Revnivtsev2011} suggested that LMXBs with
X-ray luminosities below $2\times10^{37}$ have unevolved secondary
companions while systems with higher X-ray luminosity predominantly
harbor giant donors.

In theoretical work, people usually assume that the X-ray luminosity
is directly proportional to the mass-accretion rate of compact stars
in LMXBs. The mass-accretion rate depends on the orbital period,
donor's evolution, and angular momentum loss. Donor is usually a
main sequence (MS) or a white dwarf (WD), and it may also be a red
giant (RG) or a naked helium star (He). The properties of LMXBs are
closely related to their donors. If donors are poor-hydrogen stars
(WDs or Hes), LMXBs usually are ultra-compact X-ray binaries whose
orbital periods are shorter than 80 minutes \citep{Nelson1986}. If
donors are RGs, LMXBs are called as symbiotic X-ray
binaries\citep{Masetti2006}. Up to now, there are about a dozen of
ultra-compact X-ray binaries and about 10 symbiotic X-ray binaries.

Different types of stars have different properties and evolutions.
Therefore, the donors in LMXBs determine orbital period and angular
momentum loss, which directly affects the mass-transfer rate from
the donors to NSs. Obviously, the donors really determine the
luminosity of LMXBs. Simultaneously, the observational luminosity
function of LMXBs provides important constrains for our simulating
donors' evolution.

On observations, LMXBs are divided into transient and persistent
sources.  It is difficult to estimate the X-ray luminosities of
transient LMXBs during quiescent state and outburst state. In this
paper, we focus on persistent LMXBs (PLMXBs) with accreting NS and
different donors, and investigate their properties and contributions
to total X-ray luminosity. In $\S$ 2 we present our assumptions and
describe some details of the modeling algorithm. In $\S$ 3 we
discuss the main results. In $\S$ 4 the main conclusions are given.
\section{Models}
For the simulation of binary evolution, we use rapid binary star
evolution code BSE \citep{Hurley2002} with updates by
\cite{Kiel2006}. BSE code calculates the orbital changes of binary
systems via mass variations, gravitational radiation and magnetic
braking. Details are in \S 2 of \cite{Hurley2002}.

Forming channels of LMXBs are important questions in X-ray
astronomy. There are many literatures to investigate them
\citep[e.g.,][]{Bhattacharya1991,
Podsiadlowski2002,Pfahl2003,Lin2010}. The difficulty in a
theoretical investigation involves two problems:(i)keeping the
binary bound when the massive progenitor of NS explodes in a
supernova event;(ii)a common envelope (CE) phase. They have great
effects on LMXBs. The following two subsections give descriptions.

\subsection{Common Envelope Evolution} In binary system, due to
orbital angular momentum loss or stellar expansion, a star can
overflow its Roche lobe. If the mass ratio of the components
($q=M_{{\rm donor}}/M_{{\rm accretor}}$) at the onset of Roche lobe
overflow is larger than a certain critical value $q_{\rm c}$, the
mass transfer is dynamically unstable and results in the formation
of a CE. The issue of the criterion for dynamically unstable Roche
lobe overflow, $q_{\rm c}$, is still open. Based on the polytropic
models, \cite{Webbink1988} gave $q_{\rm c}$ for red giants by
\begin{equation}
q_{\rm c}=0.362+\frac{1}{3\times(1-M_{\rm c}/M_{\rm donor})},
\label{eq:qcrit}
\end{equation}
where $M_{\rm c}$ is the core mass of donor. However, this $q_{\rm
c}$ is obtained under conservative Roche lobe overflow.
\cite{Han2001,Han2002} showed that $q_{\rm c}$ depends heavily on
the assumed mass-transfer efficiency. They found that $q_{\rm c}$
almost linearly increases with the amount of the mass and momentum
lost during mass transfer. In \cite{Han2002} the critical mass ratio
$q_{\rm c}$ is between 1.1 and 1.3. In this work, we take $q_{\rm
c}=$Eq.(\ref{eq:qcrit}), $q_{\rm c}=1.2$ and $q_{\rm c}=2.0$ in
different simulations, respectively.

Although many efforts have been devoted to understanding the
evolution of CE \citep[e.g.,][]{Ricker2008,Ge2010,Deloye2010}, the
knowledge about it is still poor. It is generally assumed that the
orbital energy of the binary is used to expel the envelope of the
donor with an efficiency $\alpha_{\rm ce}$, which is called as
$\alpha$-algorithm. \cite{Nelemans2000} suggested to describe the CE
evolution by an algorithm based on the equation for the system
orbital angular momentum balance which implicitly assumes the
conservation of energy\citep{Webbink1984}, which is called as
$\gamma$-algorithm. Following \cite{Lu2006}, for CE evolution in
different simulations we use $\alpha_{\rm ce}\lambda_{\rm ce}=1.0$
in $\alpha$-algorithm and $\gamma=1.5$ in $\gamma$-algorithm,
respectively. Here $\lambda_{\rm ce}$ is a structure parameter which
depends on the evolutionary stage of the donor.

\subsection{Formation Channels of Neutron Stars and Kick Velocity} In
X-ray binaries, NSs can be formed via three
channels\citep[e.g.,][]{Ivanova2008,Kiel2008}: (i) Core-collapse
supernovae (CCSN) for a star with main-sequence mass $ M/M_\odot
\geq 11$; (ii)Evolution induced collapse (EIC) of a helium star with
a mass between $1.4$ and $2.5 M_{\odot}$ in which the collapse is
triggered by electron capture on $^{20}$Ne and $^{24}$Mg
\citep{Miyaji1980}; (iii) Accretion-induced collapses (AIC) for an
accreting ONeMg WD whose mass reaches the Chandrasekhar limit.
Response of accreting ONeMg WD is treated in the same way as the
evolution of CO WD \citep[see details in][]{Lu2009}.

Nascent NS receives additional velocity  (``kick'') due to some
still unclear process that disrupts spherical symmetry during the
collapse or later Dichotomous nature of kicks which was suggested
quite early by \citet{Katz1975}. Observationally, the kick is not
well constrained due to numerous selection effects. Currently, high
kicks ($\sim100\,\rm km\ s^{-1}$) are associated with NS originating
from CCSN, while low kicks ($\sim10\rm km\ s^{-1}$) with NS born in
EIC and AIC \citep{Pfahl2002}.

We apply to core-collapse NS Maxwellian distribution of kick
velocity $v_{\rm k}$
\begin{equation}
P(v_{\rm k})=\sqrt{\frac{2}{\pi}}\frac{v^2_{\rm k}}{\sigma^3_{\rm
k}}e^{-v^2_{\rm k}/2\sigma^2_{\rm k}}.
\end{equation}
$\sigma_{\rm k}= 190$ and 400 km s$^{-1}$ for CCSN, while
$\sigma_{\rm k}^{*}= 20$  and 10 km s$^{-1}$ for EIC and AIC in
different simulations.

\subsection{X-ray Luminosity}
The X-ray luminosity of the accreting NS can be approximated by
\begin{equation}
\begin{array}{l}
L_{\rm bol}=\eta \dot{M}_{\rm NS}{\rm c}^2 \simeq
5.7\times10^{35}{\rm erg \
s^{-1}}(\frac{\eta}{0.1})(\frac{\dot{M}_{\rm NS}}{{10^{-10}{ M_\odot
{\rm yr^{-1}}}}})
\end{array}\label{eq:lx}
\end{equation}
where $\eta \simeq 0.1$ is the efficiency of accretion onto the NS
and $\dot{M}_{\rm NS}$ is the mass-accretion rate of the NS.
Super-Eddington accretion rates may be important in the formation of
low-mass X-ray binaries and millisecond pulsars\citep{Webbink1997}.
We assume that $\dot{M}_{\rm NS}=\min({\dot{M}_{\rm NS},\eta_{\rm
Edd}\times\dot{M}_{\rm Edd}})$, where $\dot{M}_{\rm Edd}$ is
Eddington limit given by
\begin{equation}
\dot{M}_{\rm Edd}=2.08\times10^{-3}(1+X)^{-1} R_{\rm NS} M_\odot
{\rm yr}^{-1}.
\end{equation}
Here, $X$ is the hydrogen mass fraction. $\eta_{\rm Edd}$ is the
factor to allow super-Eddington luminosities, taken to be 5
\citep{Begelman2002,zuo2011}. To transform the bolometric luminosity
into the X-ray luminosity, a bolometric correction factor $\eta_{\rm
bol}$ is introduced by $L_{\rm X}=\eta_{\rm bol}L_{\rm bol}$.
Following \cite{Belczynski2008}, we take $\eta_{\rm bol}=0.55$.

Roche overflow-fed systems are subject to a thermal disk instability
and may appear either as persistent or transient X-ray sources
depending on the mass transfer rate. A system becomes a transient
X-ray source when the mass-transfer rate falls below a certain
critical value, $\dot{M}_{\rm crit}$. For hydrogen-rich disks (The
donors are MSs or RGs ), we use the work of \cite{Paradijs1996}.
Applying to Eq.(\ref{eq:lx}), $\dot{M}_{\rm crit}$ for hydrogen-rich
disk is given by
\begin{equation}
\dot{M}_{\rm crit}=1.8\times10^{15}P_{\rm orb}^{1.07} \ \ \ {\rm
g/s}, \label{eq:hydr}
\end{equation}
where $P_{\rm orb}$ is orbital period in hours. For disks with
heavier elements, we use the work of \cite{Menou2002}:
\begin{equation}
\dot{M}_{\rm crit}=\left\{
\begin{array}{ll}
 5.9\times10^{16}M_{\rm NS}^{-0.87}R_{\rm
 d}^{2.62}\alpha^{0.44}_{0.1}\  {\rm g/s}, &{\rm He\ rich}\\
 1.2\times10^{16}M_{\rm NS}^{-0.74}R_{\rm
 d}^{2.21}\alpha^{0.42}_{0.1}\ \ \  {\rm g/s}, &{\rm C\ rich}\\
 5.0\times10^{16}M_{\rm NS}^{-0.68}R_{\rm
 d}^{2.05}\alpha^{0.45}_{0.1}\ \ \  {\rm g/s}, &{\rm O\ rich}\\
\end{array}
\right. \label{eq:nhydr}
\end{equation}
where $R_{\rm d}$ is a maximum disk radius (2/3 of accretor Roche
lobe radius) in $10^{10}$ cm,  $\alpha_{0.1}=\alpha/0.1$ in which
$\alpha=0.1$ is a viscosity parameter.

If $\dot{M}_{\rm NS}>\dot{M}_{\rm crit}$ or wind-fed accretion, the
system is a PLMXB whose X-ray luminosity is determined by Eq.
(\ref{eq:lx}). If $\dot{M}_{\rm NS}<\dot{M}_{\rm crit}$ in Roche
overflow accretion, the system is a transient source. In this work
we focus on  PLMXBs.

\section{Results}
We use Monte Carlo method to simulate the initial binaries. For
initial mass function, mass-ratios, and separations of components in
binary systems, we adopt the distributions used by us in
\cite{Lu2006,Lu2008}. We assume that all binaries have initially
circular orbits. After a supernova, new parameters of the orbit are
derived using standard formulae, \citep[e. g., ][]{Hurley2002}.
Table \ref{tab:case} lists all cases considered in the present work.
Our model is normalized to formation of one binary with $M_1 \geq
0.8 M_\odot$\ per year \citep{Yungelson1993}. We use $1\times10^7$
binary systems in our Monte-Carlo simulations.

In this work, a binary is considered as PLMXB if it satisfies the
following conditions:\\
(i)Binary includes an NS and its companion's mass is lower than 6 $M_\odot$;\\
(ii)Binary orbital period and the mass-accretion rate of NS satisfy
Eqs.(\ref{eq:hydr}) and (\ref{eq:nhydr}).\\
Here, we call both low- and intermediate-mass X-ray binaries as
LMXBs.

\begin{table}
\centering
 \begin{minipage}{170mm}
  \caption{Parameters of the models for PLMXBs' populations.
          }
  \tabcolsep1.5mm
 \begin{tabular*}{160mm}{ccccc}
 \cline{1-5}
\multicolumn{1}{c}{cases}&\multicolumn{1}{c}{$\sigma_{\rm k}$(km
s$^{-1}$)}&\multicolumn{1}{c}{CE}&\multicolumn{1}{c}{$\sigma_{\rm
k}^{*}$(km s$^{-1}$)}&\multicolumn{1}{c}{$q_{\rm
c}$}\\
 \cline{1-5}
case 1&190&$\alpha_{\rm ce}\lambda_{\rm
ce}=1.0$&20&Eq.(\ref{eq:qcrit})\\
case 2&400&$\alpha_{\rm ce}\lambda_{\rm
ce}=1.0$&20&Eq.(\ref{eq:qcrit})\\
case 3&190&$\gamma=1.5$&20&Eq.(\ref{eq:qcrit})\\
case 4&400&$\gamma=1.5$&20&Eq.(\ref{eq:qcrit})\\
case 5&190&$\alpha_{\rm ce}\lambda_{\rm
ce}=1.0$&10&Eq.(\ref{eq:qcrit})\\
case 6&190&$\alpha_{\rm ce}\lambda_{\rm
ce}=1.0$&20&$q_{\rm c}=1.2$\\
case 7&190&$\alpha_{\rm ce}\lambda_{\rm
ce}=1.0$&20&$q_{\rm c}=2.0$\\

 \cline{1-5}
 \label{tab:case}
\end{tabular*}
\end{minipage}
\end{table}

\begin{table*}
\centering
 \begin{minipage}{170mm}
  \caption{Different models of PLMXBs' population. The first column gives the model number according to Table \ref{tab:case}.
  Columns 2 to 11 give the birthrates and the numbers of PMLXBs with different
  kinds of donors, respectively.  Total birthrate and number are
  showed in columns 12 and 13, respectively. NS + MS means that accreting NS has a MS donor,  NS + RG (Roche) and NS + RG (Wind) mean that accreting NS is
  fed via Roche lobe overflow and stellar wind from a RG donor, NS + WD represents that  accreting NS has a WD donor, and
  NS + He represents that accreting NS has a naked He donor, respectively.}
  \tabcolsep0.3mm
  \begin{tabular*}{160mm}{ccccccccccccc}
  \cline{1-13}
\multicolumn{1}{c}{Cases}&
\multicolumn{2}{c}{NS+MS}&\multicolumn{2}{c}{NS+RG}
&\multicolumn{2}{c}{NS+RG}&
\multicolumn{2}{c}{NS+He}&\multicolumn{2}{c}{NS+WD}&
\multicolumn{2}{c}{Total}\\
\multicolumn{1}{c}{}& \multicolumn{2}{c}{}&\multicolumn{2}{c}{Roche}
&\multicolumn{2}{c}{Wind}&
\multicolumn{2}{c}{}&\multicolumn{2}{c}{}&
\multicolumn{2}{c}{}\\
\multicolumn{1}{c}{}&\multicolumn{1}{c}{BIR}
&\multicolumn{1}{c}{NUM}
&\multicolumn{1}{c}{BIR}&\multicolumn{1}{c}{NUM}
&\multicolumn{1}{c}{BIR}
&\multicolumn{1}{c}{NUM}&\multicolumn{1}{c}{BIR}&\multicolumn{1}{c}{NUM}&
\multicolumn{1}{c}{BIR}&\multicolumn{1}{c}{NUM}&\multicolumn{1}{c}{BIR}&\multicolumn{1}{c}{NUM}\\
(1)&(2)&(3)&(4)&(5)&(6)&(7)&(8)&(9)&(10)&(11)&(12)&(13)\\
  \cline{1-13}
\multicolumn{1}{l}{CCSN}&\multicolumn{12}{c}{}\\

case 1 & 4.1$\times
10^{-5}$&68000&$1.1\times10^{-5}$&830&$4.6\times10^{-6}$&250&$3.7\times10^{-5}$&110&$1.1\times10^{-5}$&50&$1.1\times10^{-4}$&70000\\
case 2 & 1.5$\times
10^{-5}$&23000&$7.4\times10^{-6}$&460&$2.7\times10^{-7}$&$<10$&$4.2\times10^{-6}$&20&$3.7\times10^{-6}$&30&$2.5\times10^{-5}$&24000\\
case 3 & 7.0$\times
10^{-5}$&100000&$4.1\times10^{-5}$&2000&$6.4\times10^{-6}$&140&$1.7\times10^{-5}$&50&$9.3\times10^{-6}$&50&$1.3\times10^{-4}$&110000\\
case 4 & 5.2$\times
10^{-5}$&78000&$3.0\times10^{-5}$&1200&$2.0\times10^{-6}$&30&$3.6\times10^{-6}$&10&$1.6\times10^{-6}$&10&$8.2\times10^{-5}$&80000\\
case 5 & 4.1$\times
10^{-5}$&68000&$1.1\times10^{-5}$&830&$4.6\times10^{-6}$&250&$3.7\times10^{-5}$&110&$1.1\times10^{-5}$&50&$1.1\times10^{-4}$&70000\\
case 6 & 4.1$\times
10^{-5}$&68000&$1.2\times10^{-5}$&840&$4.6\times10^{-6}$&250&$3.6\times10^{-5}$&90&$9.5\times10^{-6}$&40&$1.4\times10^{-4}$&70000\\
case 7 & 4.0$\times
10^{-5}$&66000&$1.5\times10^{-5}$&850&$4.6\times10^{-6}$&250&$3.4\times10^{-5}$&70&$6.1\times10^{-6}$&30&$1.1\times10^{-4}$&68000\\

 \cline{1-13}\multicolumn{1}{l}{AIC}&\multicolumn{12}{c}{}\\

case 1 & 2.1$\times
10^{-6}$&180&$1.1\times10^{-5}$&300&---&---&$3.4\times10^{-4}$&1700&$2.0\times10^{-4}$&470&$4.7\times10^{-4}$&2700\\
case 2 & 2.1$\times
10^{-6}$&180&$1.1\times10^{-5}$&300&---&---&$3.4\times10^{-4}$&1700&$1.9\times10^{-4}$&450&$4.6\times10^{-4}$&2600\\
case 3 & 4.2$\times
10^{-6}$&470&$2.0\times10^{-5}$&340&---&---&$6.2\times10^{-5}$&280&$6.9\times10^{-5}$&280&$1.3\times10^{-4}$&1400\\
case 4 & 4.2$\times
10^{-6}$&460&$2.0\times10^{-5}$&340&---&---&$6.2\times10^{-5}$&280&$6.9\times10^{-5}$&300&$1.3\times10^{-4}$&1400\\
case 5 & 2.1$\times
10^{-6}$&220&$1.1\times10^{-5}$&280&---&---&$3.4\times10^{-4}$&1700&$2.0\times10^{-4}$&490&$4.7\times10^{-4}$&2700\\
case 6 & 2.7$\times
10^{-6}$&180&$2.5\times10^{-5}$&320&---&---&$2.8\times10^{-4}$&1300&$1.5\times10^{-4}$&420&$3.9\times10^{-4}$&2200\\
case 7 & 2.7$\times
10^{-6}$&180&$9.4\times10^{-5}$&500&---&---&$1.3\times10^{-4}$&530&$8.6\times10^{-4}$&300&$2.8\times10^{-4}$&1500\\

\cline{1-13}\multicolumn{1}{l}{EIC}&\multicolumn{12}{c}{}\\

case 1 & 1.5$\times
10^{-6}$&1100&$2.9\times10^{-5}$&140&$4.2\times10^{-6}$&$<10$&$1.1\times10^{-4}$&440&$2.8\times10^{-5}$&220&$1.4\times10^{-4}$&1900\\
case 2 & 1.5$\times
10^{-6}$&1100&$2.9\times10^{-5}$&140&$4.2\times10^{-6}$&$<10$&$1.1\times10^{-4}$&440&$2.8\times10^{-5}$&220&$1.4\times10^{-4}$&1900\\
case 3 & 3.5$\times
10^{-7}$&360&$2.2\times10^{-5}$&40&$7.1\times10^{-6}$&$<10$&$6.6\times10^{-5}$&460&$4.9\times10^{-5}$&350&$1.2\times10^{-4}$&1200\\
case 4 & 3.5$\times
10^{-7}$&360&$2.2\times10^{-5}$&40&$7.1\times10^{-6}$&$<10$&$6.6\times10^{-5}$&460&$4.9\times10^{-5}$&350&$1.2\times10^{-4}$&1200\\
case 5 & 7.5$\times
10^{-7}$&850&$2.6\times10^{-5}$&90&$4.5\times10^{-6}$&$<10$&$9.4\times10^{-5}$&450&$3.0\times10^{-5}$&210&$1.4\times10^{-4}$&1600\\
case 6 & 1.5$\times
10^{-6}$&1100&$2.9\times10^{-5}$&140&$4.2\times10^{-6}$&$<10$&$1.0\times10^{-4}$&440&$2.7\times10^{-5}$&210&$1.7\times10^{-4}$&1900\\
case 7 & 1.5$\times
10^{-6}$&1100&$2.9\times10^{-5}$&150&$4.1\times10^{-6}$&$<10$&$8.9\times10^{-5}$&310&$2.3\times10^{-5}$&190&$1.8\times10^{-4}$&1800\\

\cline{1-13} \label{tab:result}
\end{tabular*}
\end{minipage}
\end{table*}

\subsection{Birthrates and Numbers of PLMXBs' Populations}
\label{sec:num} In our simulations, there are $\sim$ 29000 (case 2)
--- 110000 (case 3) PLMXBs in the Galaxy, and their birthrates are
$\sim$ 3.4---7.2 $\times 10^{-4}$ yr$^{-1}$. However, the number of
all observed LMXBs is less than 200 \citep{Liu2006}.
\cite{Pfahl2003} investigated LMXBs via CCSN, and obtained
birthrates for LMXBs of $10^{-6}$---$10^{-4}$ yr$^{-1}$ and
400---70000 LMXBs, a factor 10---1000 times higher than observed
number. Many authors suspected that the mismatch between the
observed number and theoretically predicted number could be related
to irradiation effects \citep[et al.][]{Hameury1993,Hurley2010}. In
this work, we do not consider LMXBs in low states driven by
irradiation-driven limit cycles. However, we still encounter the
known problem of ¡°overproduction¡± of LMXBs.

As Table \ref{tab:result} shows, NSs with different kinds of donors
in PLMXBs have different formation channels. More than 90\% of
PLMXBs have undergone CCSN, especially, for PLMXBs with
hydrogen-rich donors. However, more than 90\% of PLMXBs with He
donors or WD donors have undergone AIC and EIC.  In cases 1 and 2,
parameter $\sigma_{\rm k}$ is increased from 190 to 400 km s$^{-1}$.
The larger $\sigma_{\rm k}$ is, the more difficultly binary survive
after CCSN. Therefore, the birthrate and number in case 2 are about
1/3 of these in case 1, and there is not PLMXBs with RG (Wind)
donors. In cases 1 and 3, different algorithms of CE are used.
Usually, binary orbit after CE shortens up to $\sim$1\% of initial
one under $\alpha$-algorithm assumption, while it approximately
remains unchanged under $\gamma$-algorithm assumption. Many binaries
avoid to merge when they are undergoing CE evolution in case 3.
There are more PLMXBs in case 3 than those in case 1. Parameter
$\sigma_{\rm k}^{*}$ is decreased from 20 to 10 km s$^{-1}$ in case
5. AIC and EIC with $\sigma_{\rm k}^*=20$ km s$^{-1}$ produce wider
orbital periods than those with $\sigma_{\rm k}^*=10$ km s$^{-1}$
for the progenitors of PLMXBs although some binary systems can be
disturbed. Wider orbital periods provide enough separations so that
the secondaries can evolve to RG, and can survive after CE
evolution. Therefore, PLMXBs via AIC and EIC in case 5 are lower
than those in case 1. We also carried out a test in which
$\sigma_{\rm k}^*=190$ km s$^{-1}$, and found that the number and
birthrate of PLMXBs via AIC and EIC greatly decrease. Compared cases
6 and 7 with case 1, parameter $q_{\rm c}$ has a weak effect on
PLMXBs populations. Therefore, from this later, we only discuss
cases 1, 2, 3 and 5 to illustrate the effects of input parameters on
PLMXBs, respectively.

\begin{figure}
\includegraphics[totalheight=3.3in,width=3.3in,angle=-90]{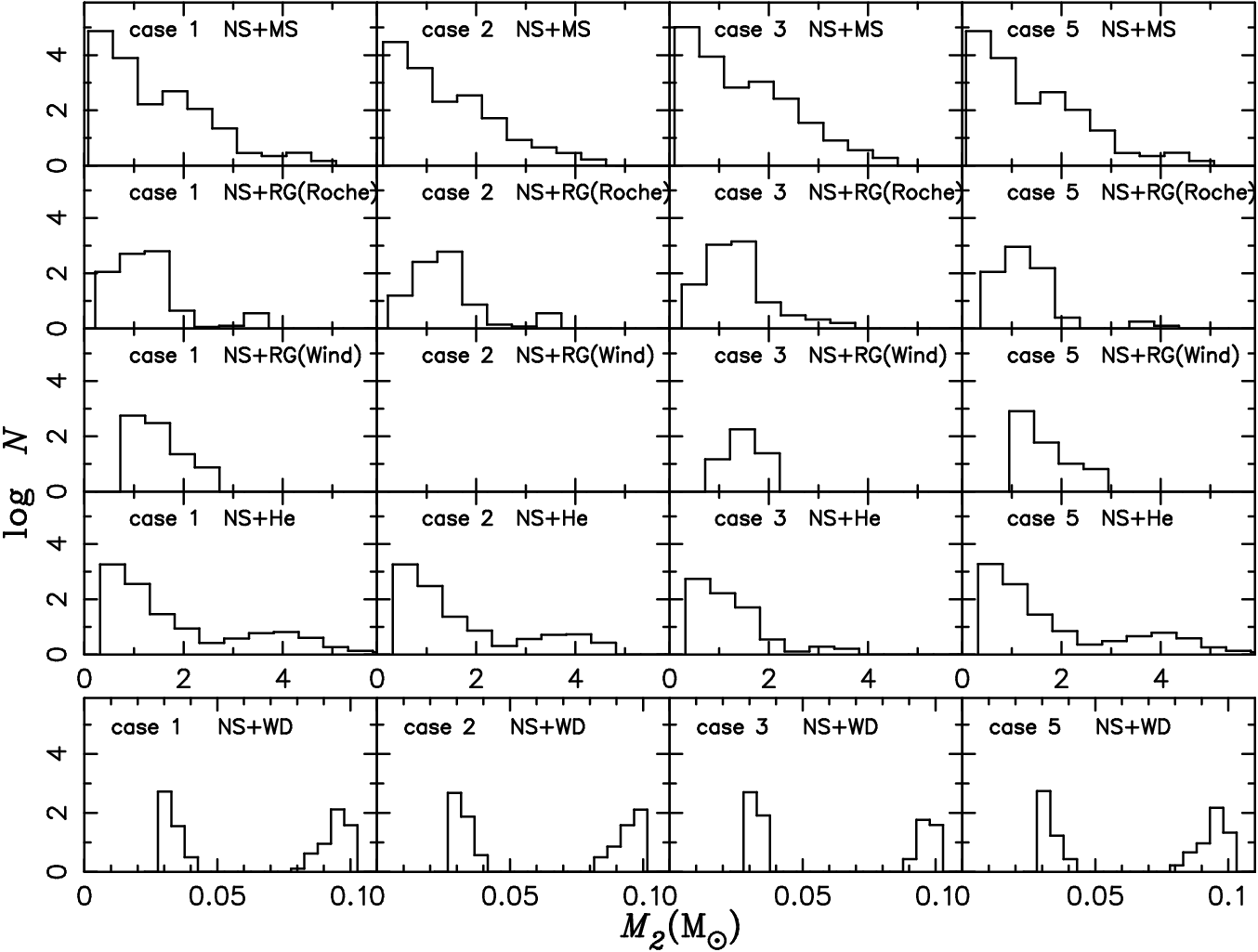}
\caption{---Number distribution of the masses of different donors in
PLMXBs. The width of the bin is 0.005 $M_\odot$ for PLMXBs with WD
donors, and they are 0.5 $M_\odot$ for others. }\label{fig:m2}
\end{figure}

The number of PLMXBs mainly depends on the masses and the evolution
phase of donors. Figure \ref{fig:m2} shows the distribution of
donors' masses. Majority of MS donors have lower masses than 1
$M_\odot$, while most of RG (Wind) donors have masses than 1
$M_\odot$ and the durations of RGs  are very shorter than those of
MSs. Therefore, majority of PLMXBs are NS + MS systems, and PLMXBs
with RG (Wind) donors are very rare.

In PLMXBs with WD donors, matter transfer is driven by gravitational
radiation. \cite{Deloye2003} and \cite{Bildsten2004} showed the
relation among WD masses, orbital periods and mass-transfer rates in
ultra-compact X-ray binaries with WD donors. According to their
results, the higher WD donors' masses are, the higher mass-transfer
rates are. The duration of PLMXBs with massive WDs is very short. As
Figure \ref{fig:m2} shows, there are two peaks in the distributions
of WDs' masses. The left peak is at $\sim$ 0.03 $M_\odot$, and the
right peak is at $\sim$ 0.09 $M_\odot$. The former mainly comes from
the PLMXBs which will translate from persistent to transient state
because the donors in these PLMXBs have low mass-loss rates which
results in long durations. The later mainly results from the PLMXBs
who have undergone the AIC. Compared with the PLMXBs around the left
peak, these PLMXBs around the right peak have short orbital periods
and high X-ray luminosities. There are $\sim$ 600
--- 900 PLMXBs with WD donors. However, the duration of LMXBs with
WD donors whose masses are lower than those in PLMXBs are very long
($\sim 10^9$ yr) because of a low mass-transfer rate ($\sim 10^{-12}
M_\odot$ yr$^{-1}$). Therefore, most of LMXBs with WD donors are
transient.

Not like WDs, naked He stars are convective. The mass transfer in
PLMXBs with He donors is driven by gravitational radiation and
magnetic braking. In general, the later is dominated
\citep{Hurley2002,Postnov2005}, and drives a mass-transfer rate of
$\sim$ $10^{-7} M_\odot$ yr$^{-1}$ in our work. Majority of He
donors' masses are between $\sim$ 0.3 --- 2.0 $M_\odot$ (Figure
\ref{fig:m2}). Therefore, there are several thousand PLMXBs with He
donors in the Galaxy. According to Table \ref{tab:result}, the
number of PLMXBs with He donors is much larger than PLMXBs with WD
donors. From the properties of type I X-ray bursts, \cite{Zand2005}
suggested that in most ultra-compact X-ray binaries the matter
accumulated on NSs is helium. This is consistent with our results
although we do not discuss transient LMXBs. In the PLMXBs plotted in
Figure \ref{fig:mapo}, there are two ultra-compact X-ray binaries
(4U 1626-67 and 4U 0614+09) which have very evolved He
donors\citep{Nelemans2010}.

\subsection{Properties of PLMXBs with Different Donors}
\begin{figure*}
\includegraphics[totalheight=6.3in,width=6.3in,angle=-90]{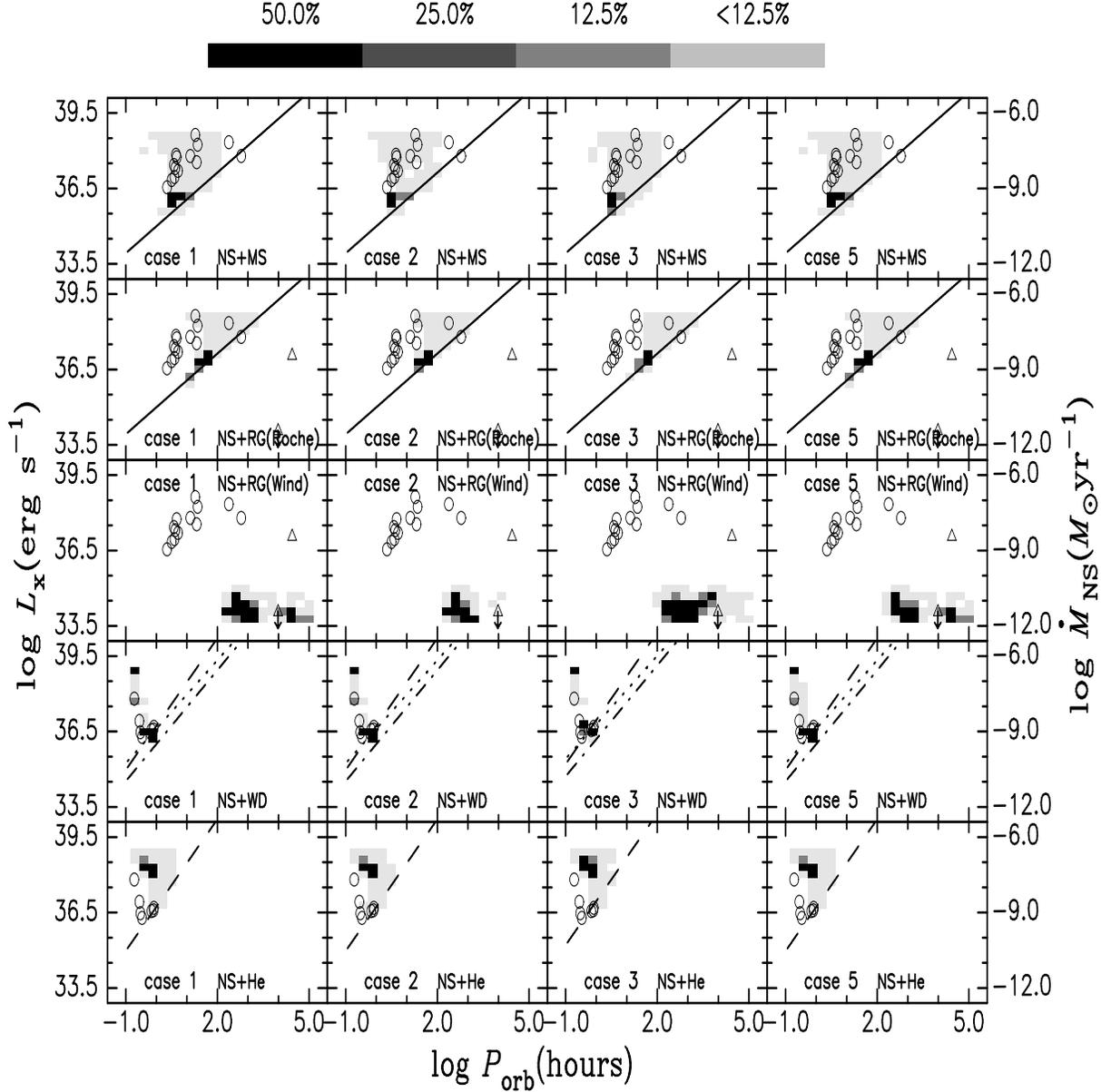}
\caption{---Distributions of the orbital periods
            vs. the X-ray luminosities or the mass-accretion rate
            $\dot{M}_{\rm NS}$ in PLMXBs.
            The gradations of gray-scale correspond to the regions where the number density of systems is,
            respectively,  within 1 -- 1/2, 1/2 -- 1/4, 1/4 -- 1/8, 1/8 -- 0 of the maximum of
            ${{\partial^2{N}}\over{\partial {\log P_{\rm orb}}{\partial {\log L_{\rm x}}}}}$. Number in every panel
            is normalized to 1.
            Circles represents the brightest PLMXBS
            from \citet{Revnivtsev2011}, and triangles represent the symbiotic X-ray binaries in
            \citet{Masetti2007}. Solids represent the relation for hydrogen-rich disk of Eq.(\ref{eq:hydr}), and dashed,
            dotted and dash-dotted lines represent the relations for He-rich, C-rich and O-rich disk of
             Eq.(\ref{eq:nhydr}), respectively.}\label{fig:mapo}
\end{figure*}

As \S \ref{sec:introduction} mentions, the donors of LMXBs basically
determine the orbital periods and mass-transfer rates which give the
X-ray luminosity. Figure \ref{fig:mapo} shows the distributions of
the orbital periods and the X-ray luminosities or the mass-accretion
rate  $\dot{M}_{\rm NS}$. PLMXBs with different donors have
different positions in Figure \ref{fig:mapo}.

\begin{figure*}
\includegraphics[totalheight=6.3in,width=6.3in,angle=-90]{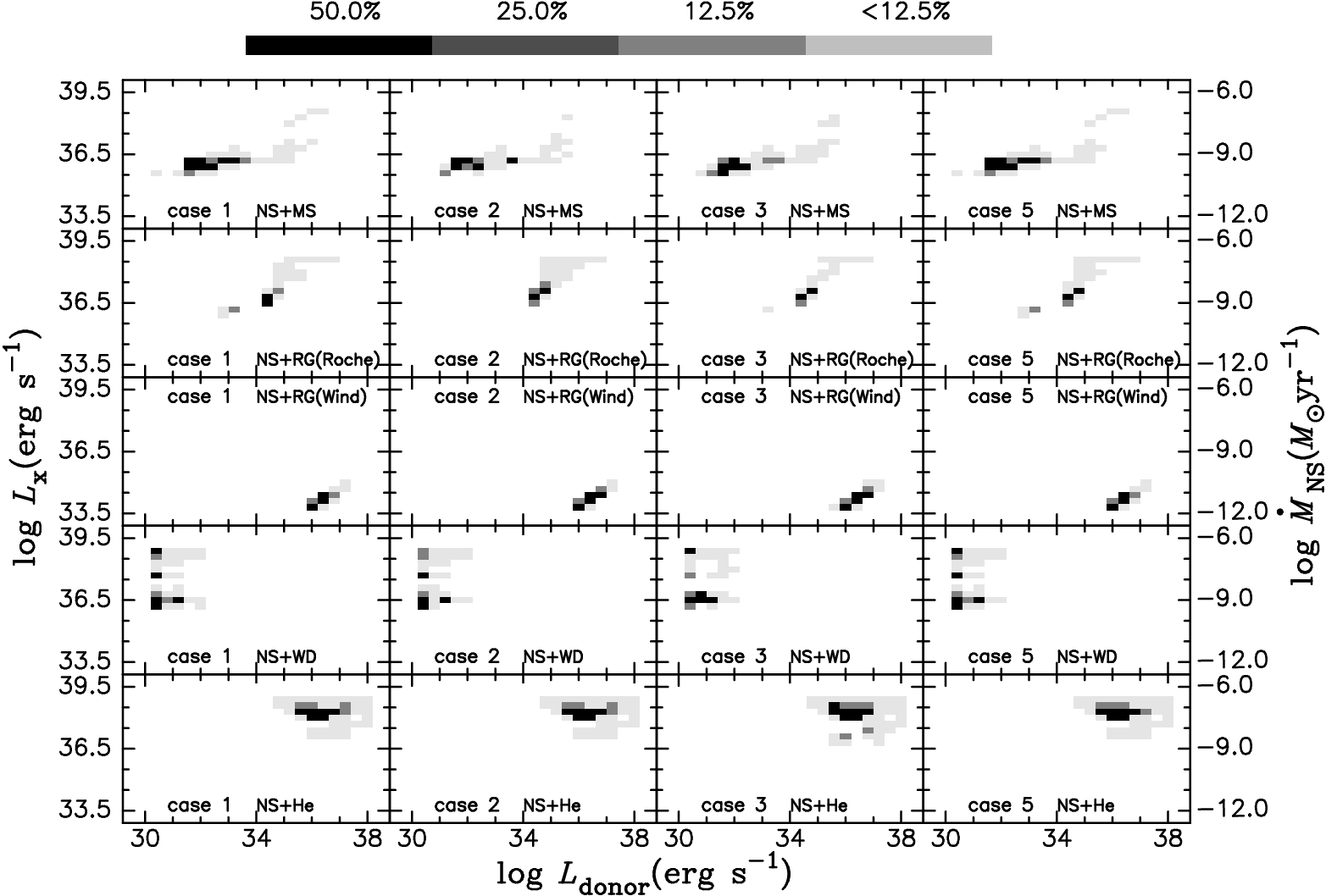}
\caption{---Similar with Figure \ref{fig:mapo}, but for
            distributions of donors' luminosity
            vs. the X-ray luminosities or the mass-accretion rate
            $\dot{M}_{\rm NS}$ in PLMXBs.}\label{fig:lxl2}
\end{figure*}

\cite{Revnivtsev2011} investigated the brightest Galactic PLMXBs
which are plotted in Figure \ref{fig:mapo}. They concluded that the
majority of PLMXBs with X-ray luminosities below $\sim
2\times10^{37}$ erg s$^{-1}$ have unevolved MS, while PLMXBs with
higher X-ray luminosity predominantly harbor giant donors. In the
panels of NS + MS in Figure \ref{fig:mapo}, $\sim$ 90\% of PLMXBs
with MS donors lie in the shallow region with a X-ray luminosity of
$\sim 10^{36}$ erg s$^{-1}$. The irradiation of LMXBs can drive mass
transfer \citep{Podsiadlowski1991,Buning2004}. In this work, we do
not consider the effect of irradiation. Therefore, compared with the
luminosities of known PLMXBs, we may underestimate the luminosity of
PLMXBs with MS donors.  In the panels of NS + RG (Roche) in Figure
\ref{fig:mapo}, our sample covers the positions of Cyg X-2 and GX
13+1 which have long orbital periods. These PLMXBs can only be
explained by NS + RG (Roche) systems in our simulations.
\cite{Orosz1999} gave good measurements for Cyg X-2 and have derived
the donor's mass around 0.6 $M_\odot$. \cite{Podsiadlowski2000}
suggested that the donor in Cyg X-2 has a mass of around 0.5
$M_\odot$ with a non-degenerated helium core and is burning hydrogen
in a shell. That is, the donor in Cyg X-2 is a sub-giant which has
undergone the violent mass loss. This is consistent with ours.
\cite{Bandyopadhyay1999} gave that the spectrum of donor in GX 13+1
clearly shows the features of the K5 III giant. Figure
\ref{fig:lxl2} gives the distributions of donors' luminosities and
mass-transfer rates determine the X-ray luminosity. As the panels of
NS + RG (Roche) and NS + RG (Wind) in Figure \ref{fig:lxl2} show,
the donors' luminosities in PLMXBs with RG (Roche) donors are much
lower than those in PLMXBs with RG (Wind), while the X-ray
luminosities in the former are much higher than those in the later.
Therefore, it is more difficult to observe the donors' luminosities
in PLMXBs with RG (Roche) donors. We suggested that donors in Cyg
X-2 and GX 13+1 are giants which fill up Roche lobes. The orbital
periods and X-ray luminosities of 2 known symbiotic X-ray binaries
(GX 1+4 and 4U 1700+24) are measured, which is plotted by triangles
in Figure \ref{fig:mapo}. NS + RG (Wind) systems can cover 4U
1700+24 very well, but our models can not explain GX 1+4. A detailed
investigation of symbiotic X-ray binaries is being carried out
(L\"{u} et al. in preparation).

In our simulations, the orbital periods of PLMXBs with WD or He
donors are shorter than 80 minutes, and they are ultra-compact X-ray
binaries. About 10\% --- 34\% of WDs in NS + WD systems are He WDs,
and 66\% --- 90\% are CO WDs. NS + ONeMg WD systems are negligible.
As Figure \ref{fig:mapo} shows, PLMXBs with WD donors are agree with
observations, while PLMXBs with He donors have X-ray luminosities
higher than those of observed ultra-compact X-ray binaries.
\cite{Nelemans2010} suggested that two ultra-compact X-ray binaries
4U 1626-67 and 4U 0614+09 have very evolved He donors, and their
X-ray luminosities are $3.2\times10^{36}$ erg s$^{-1}$ and
$3.4\times10^{36}$ erg s$^{-1}$. In our model, the mass transfer in
PLMXBs with He donors is driven by the magnetic braking, which
produces the X-ray luminosity of $\sim 10^{38}$  erg s$^{-1}$. We
may overestimate the work efficiency of the magnetic braking driving
mass transfer. As panels of NS + WD and NS + He in Figure
\ref{fig:lxl2} show, the donors' luminosities in PLMXBs with He
donors are much higher than those in PLMXBs with WD donors. This
difference may  be a way via which we can distinguish He donors from
WD donors.

\begin{figure}
\includegraphics[totalheight=3.3in,width=3.3in,angle=-90]{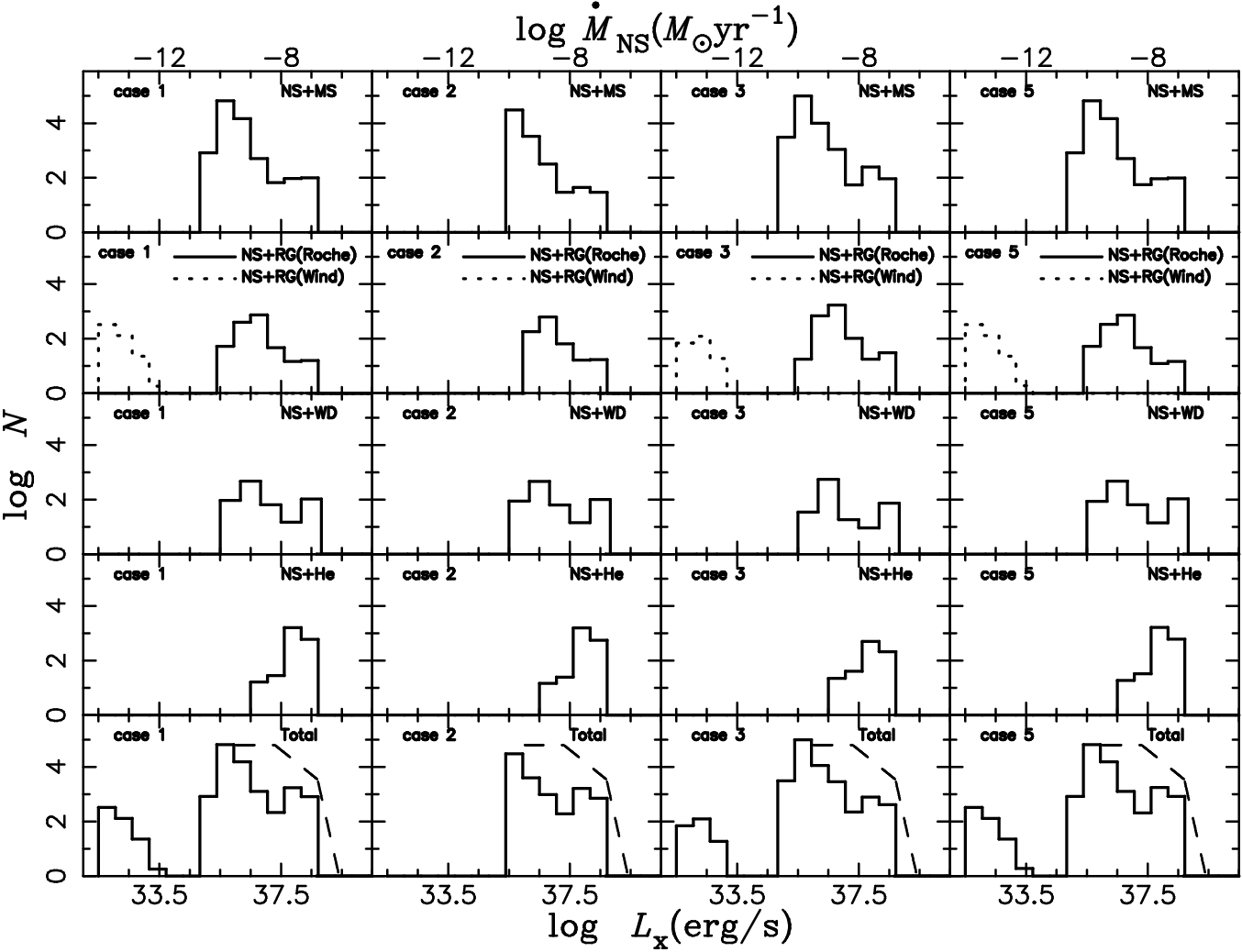}
\caption{---Number distribution of the X-ray luminosities
(mass-accretion rates) of NSs in PLMXBs with different donors. The
width of the bin for $\log L_{\rm x} {\rm (erg/s)}$ is 0.5. The
dashed lines in the total panels represent the power-law X-ray
luminosity function for LMXBs in \citet{Gilfanov2004}.
}\label{fig:lxg}
\end{figure}

Figure \ref{fig:lxg} gives the distribution of the X-ray
luminosities (mass-accretion rates) of NSs in PLMXBs with different
donors. The X-ray luminosities between $\sim 6\times 10^{32}$ ---
$6\times 10^{33}$ erg s$^{-1}$ mainly come from PLMXBs with RG
(Wind). The X-ray luminosities between $\sim 10^{35}$ --- $10^{37}$
erg s$^{-1}$ originate from PLMXBs with MS, RG (Roche) and WD
donors, in which PLMXBs with MS are dominated. The X-ray
luminosities between $\sim 10^{37}$ --- $10^{39}$ erg s$^{-1}$
mainly originate from PLMXBs with He donors in which the mass
transfer is driven by magnetic braking. \cite{Postnov2005} suggested
that observed X-ray luminosity function ($>2\times10^{37}$ erg
s$^{-1}$ ) of LMXBs can generally be explained by the accretion of
matter onto a NS with magnetic stellar wind, which agrees with our
results. However, the observed X-ray luminosity function ( up to
$\sim2\times10^{37}$ erg s$^{-1}$ ) of LMXBs can be explained by
PLMXBs with MS and RG (Roche) donors. In BSE code, gravitational
radiation is only efficient for binaries with orbital periods less
than 3 hours.  In order to explain a gap of CVs between 2 and 3 hr
in the otherwise smooth period-mass distribution, BSE code does not
apply magnetic braking when the primary is a fully convective MS
star whose mass is lower than $0.35 M_\odot$. There is not the
magnetic braking in NS + MS systems if MS's mass is larger than $1.2
M_\odot$ because it has no convective envelope. In our work, less
than 30\% of PLMXBs with MS donors  have a donor whose mass is
between $0.35 M_\odot$ and $1.2 M_\odot$. Therefore, the mass
transfer in PLMXBs with X-ray luminosity function ( up to
$\sim2\times10^{37}$ erg s$^{-1}$ ) is mainly driven by stellar
evolution or magnetic braking, but not gravitational radiation.

In the bottom panels of Figure \ref{fig:lxg}, we give the
distributions of total X-ray luminosities which are different from
the power-law X-ray luminosity function (showed by dashed line)
observed by \citet{Gilfanov2004}. There are two main reasons: i)Our
work does not include the transient X-ray luminosity. The transient
systems in outburst reach high (close to Eddington) X-ray
luminosities. \cite{Belczynski2008} suggested that the X-ray
luminosities in outburst are between $\sim 10^{37}$ and $10^{38}$
erg s$^{-1}$. ii)We overestimate the X-ray luminosities of PLMXBs
with He donors. Of course, we must obtain total population of LMXBs
including transient and persistent systems with NSs and black hole
before we compare X-ray luminosity function with observational one.
In the future, we will try doing it.

\section{Conclusion}
We perform a detailed study of donors in PLMXBs, employing the
population synthesis approach to the evolution of binaries. We
estimate that there are $\sim$ 29000
--- 110000 PLMXBs in the Galaxy, and their birthrates are
$\sim$ 3.4---7.2 $\times 10^{-4}$ yr$^{-1}$. PLMXBs with different
donors have different formation channels. Our numerical simulation
shows that more than 90\% of PLMXBs have MS donors, and PLMXBs with
RG (Wind) donors are negligible. Most of NSs in PLMXBs with
hydrogen-rich donors have undergone CCSN, while more than 90\% of
NSs in PLMXBs with He donors or WD donors have undergone AIC and
EIC.

In PLMXBs with MS donors, the orbital periods are between $\sim$ 1
hour and 100 hours, and the mass transfer is driven by donor
evolution or magnetic braking. Our population synthesis code shows
that their X-ray luminosities mainly are around $\sim 10^{36}$ erg
s$^{-1}$. Similarly, in PLMXBs with RG donors via Roche lobe
overflow(Roche), the mass transfer is driven by donor evolution, but
orbital periods are between $\sim$ 10 hours and 1000 hours. Their
X-ray luminosities are around $\sim 10^{37}$ erg s$^{-1}$. The 2
known LMXBs ( Cyg X-2 and GX 13+1) can belong to PLMXBs with RG
(Roche). PLMXBs with RG (Wind) donors have the longest orbital
periods and low X-ray luminosities. Their contributions to X-ray
luminosities can be negligible. In PLMXBs with He donors, the
orbital periods are shorter than 80 minutes, and the mass transfer
is mainly driven by magnetic braking. Results of our numerical
simulations predict that PLMXBs with X-ray luminosities around $\sim
10^{38}$ erg s$^{-1}$ mainly come from binaries with He donors, but
their X-ray luminosities may be overestimated in our work. In PLMXBs
with WD donors, the orbital periods are shorter than 1 hour, and the
mass transfer is mainly driven by gravitational radiation. According
to results of our population synthesis code, their X-ray
luminosities are between $\sim 6\times 10^{35}$---$10^{39}$ erg
s$^{-1}$, and most of LMXBs with WD donors are transient.

In this work we do not consider transient LMXBs with different
donors. In further work, we will investigate donors in persistent
and transient LMXBs, and discuss X-ray luminosity function of LMXBs.

\section*{Acknowledgments}
We thank an anonymous referee for his/her comments which helped to
improve this paper. This work was supported by the National Natural
Science Foundation of China under No.11063002, the Knowledge
Innovation Program of the Chinese Academy of Sciences (Grant No.
KJCX2-YW-T09), National Basic Research Program of China (973 Program
2009CB824800), Natural Science Foundation of Xinjiang under
Nos.2009211B01 and 2010211B05, Foundation of Huoyingdong under
No.121107, Foundation of Ministry of Education under No.211198, and
Doctor Foundation of Xinjiang University (BS100106).

\bibliographystyle{apj}
\bibliography{lglmn}
\label{lastpage}
\end{document}